\documentclass[fleqn,usenatbib]{mnras}

\usepackage{newtxtext,newtxmath}
\usepackage{hyperref}
\usepackage{lineno}
\usepackage{tablefootnote}
\usepackage[T1]{fontenc}

\DeclareRobustCommand{\VAN}[3]{#2}
\let\VANthebibliography\thebibliography
\def\thebibliography{\DeclareRobustCommand{\VAN}[3]{##3}\VANthebibliography}

\usepackage{graphicx}	
\usepackage{amsmath}	

\title[Iberian superbolide independently observed]{The 18 May 2024 Iberian superbolide from a sunskirting orbit: USG space sensors and ground-based independent observations} 

\author[Peña-Asensio et al.]{
E. Peña-Asensio,$^{1,2}$\thanks{E-mail: eloy.pena@polimi.it, eloy.peas@gmail.com}
P. Grèbol-Tomàs,$^{2,3}$
J. M. Trigo-Rodríguez,$^{2,3}$
P. Ramírez-Moreta,$^{4}$
and R. Kresken$^{4}$
\\
$^{1}$Department of Aerospace Science and Technology, Politecnico di Milano, Via La Masa 34, 20156 Milano, Italy\\
$^{2}$Institut de Ciències de l’Espai (ICE, CSIC) Campus UAB, C/ de Can Magrans s/n, 08193 Cerdanyola del Vallès, Catalonia, Spain\\
$^{3}$Institut d’Estudis Espacials de Catalunya (IEEC) 08034 Barcelona, Catalonia, Spain\\
$^{4}$ESA/ESAC/Planetary Defence Office (OPS-SP) Camino Bajo del Castillo s/n, 28692 Villanueva de la Cañada, Madrid, Spain
}

\date{Accepted XXX. Received YYY; in original form May 23, 2024}

\pubyear{\the\year{}}

\begin{document}
\label{firstpage}
\pagerange{\pageref{firstpage}--\pageref{lastpage}}
\maketitle

\begin{abstract} On 18 May 2024, a superbolide traversed the western part of the Iberian Peninsula, culminating its flight over the Atlantic Ocean and generating significant media attention. This event was caused by a weak carbonaceous meteoroid of 1 m, entering the atmosphere at 40.4 km\,s$^{-1}$ with an average slope of 8.5$^\circ$. The luminous phase started at 133 km and ended at an altitude of 54 km. The meteoroid's heliocentric orbit had an inclination of 16.4$^\circ$, a high eccentricity of 0.952, a semi-major axis of 2.4 au, and a short perihelion distance of 0.12 au. The superbolide was recorded by multiple ground-based stations of the Spanish Fireball and Meteorite Network (SPMN) and the European Space Agency (ESA), as well as by the U.S. Government (USG) sensors from space. Due to the absence of observable deceleration, we successfully reconciled satellite radiometric data with a purely dynamic atmospheric flight model, constraining the meteoroid's mass and coherently fitting its velocity profile. Our analysis shows a good agreement with the radiant and velocity data reported by the Center for Near-Earth Object Studies (CNEOS), with a deviation of 0.56$^\circ$ and 0.1 km\,s$^{-1}$, respectively. The presence of detached fragments in the lower part of the luminous trajectory suggests that the meteoroid was a polymict carbonaceous chondrite, containing higher-strength macroscopic particles in its interior due to collisional gardening, or a thermally processed C-type asteroid. The orbital elements indicate that the most likely source is the Jupiter-Family Comet (JFC) region, aligning with the SOHO comet family, as its sunskirting orbit is decoupled from Jupiter. This event provides important information to characterize the disruption mechanism of near-Sun objects.


\end{abstract}

\begin{keywords}
meteorites, meteors, meteoroids -- comets: general -- minor planets, asteroids: general
\end{keywords}



\section{Introduction}

On 18 May 2024, an exceptionally luminous fireball was observed over Spain and Portugal. This event was captured on video by casual observers and quickly disseminated through the media. Additionally, it was recorded by various wide-field and multi-camera stations that continuously monitor the sky over the Iberian Peninsula. The U.S. Government (USG) sensors also detected the event from space, as reported on the Center for Near-Earth Object Studies (CNEOS) fireball website\footnote{\url{https://cneos.jpl.nasa.gov/}}. Following confirmation of its detection from space\footnote{It was also detected by ESA's Meteosat 3Gen (MTG) satellite via its Lightning Imager instrument: \url{https://www.esa.int/Applications/Observing_the_Earth/Meteorological_missions/meteosat_third_generation/Fireball_witnessed_by_weather_satellite}}, the event can be formally classified as a superbolide \citep{1999md98.conf...37C}. These exceptionally bright fireballs are produced by the hypersonic atmospheric entry of meter-sized natural projectiles \citep{Ceplecha1998SSRv, Silber2018AdSpR}. The study of superbolides provides valuable insights into the physical properties, dynamics, and impact hazard issues associated with the near-Earth object population \citep{2017JIMO...45...91K, Trigo2022}. 

The CNEOS database, managed by NASA's Jet Propulsion Laboratory, archives data on fireball events detected via various USG satellite sensors \citep{1994hdtc.conf..199T}. Superbolides are relatively rare events that can occur over remote areas. USG sensors provide near-global coverage of large meteoroids and small asteroids impacting the Earth's atmosphere, unlike the ground-based meteor network which only monitors a relatively small, fixed atmospheric volume. Therefore, the CNEOS database is of scientific interest as it extends the projectile flux estimations by utilizing the entire planet as a detector, capturing events that are typically singular occurrences for other techniques \citep{2002Natur.420..294B}. As of May 2024, the database contains 979 fireball events, with velocity vector and position data available for 310 cases. For these events, detailed information is provided, including geographic coordinates and altitude of the radiated energy peak, vector velocity components in an Earth-Centered, Earth-Fixed (ECEF) reference frame, and radiated and impact energies. These energies, linked by an empirical formula, are among the most robust parameters provided by CNEOS \citep{2002Natur.420..294B}. 

However, specific information about these sensors remains undisclosed as they are classified data. In any case, we previously prepared our \textit{3D-FireTOC} software to compute future detections based on CNEOS-released data \citep{2021MNRAS.504.4829P} and performed an analysis of the CNEOS database \citep{2022AJ....164...76P}. The accuracy of the CNEOS data has been assessed through comparisons with ground-based observations of fireballs \citep{2019MNRAS.483.5166D, 2022AJ....164...76P, 2023ApJ...953..167B, 2024Icar..40815844P}. To date, only 17 fireballs in the CNEOS database have been benchmarked with counterparts. Here we report and compare a new event detected by the USG sensors and multiple ground-based networks. 



\section{Observations and methods}

\subsection{Ground-based observation}

\begin{figure*}
	\includegraphics[width=1.8\columnwidth]{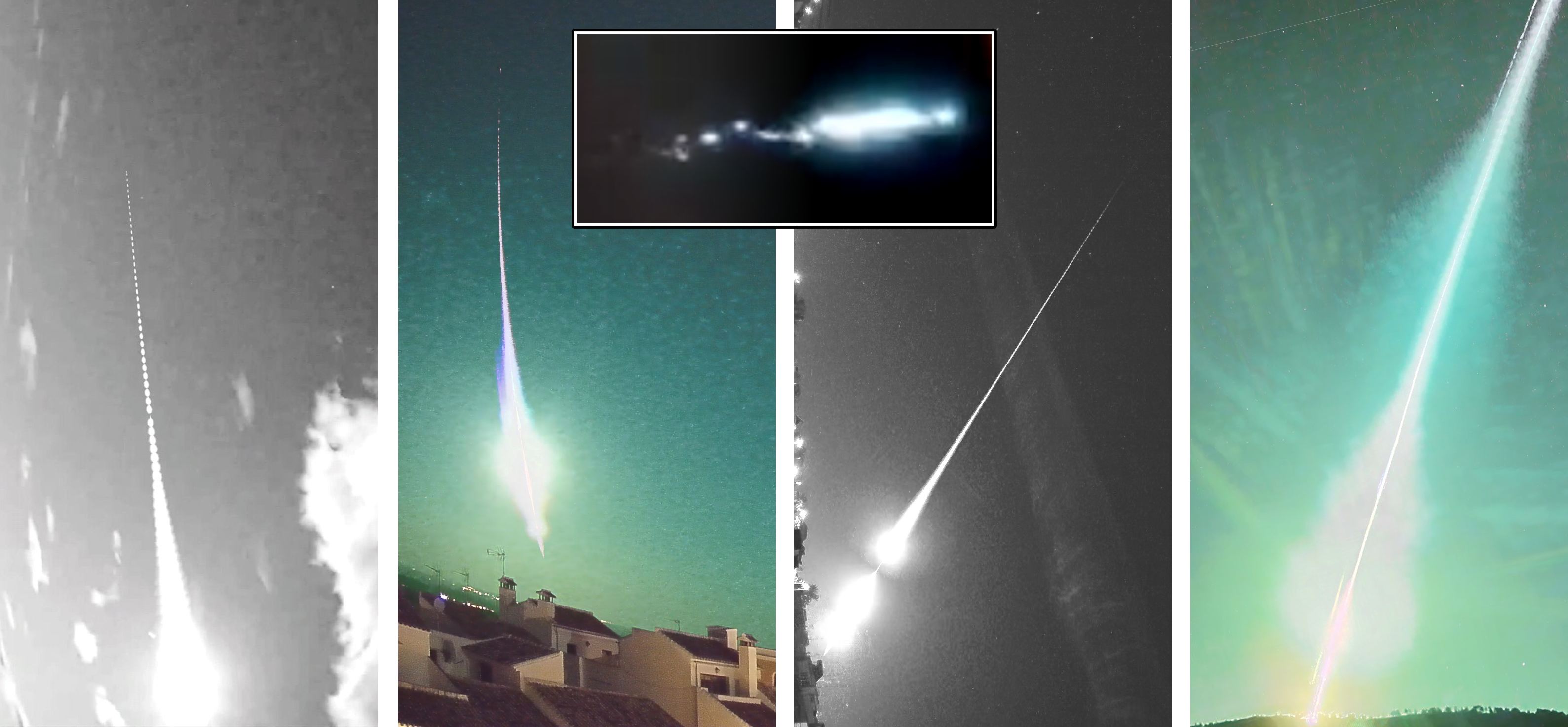}
    \caption{The four videos used in this work to study the SPMN180524F superbolide. The images are max-combined from the video frames. Some saturated frames have been removed for illustration purposes. From left to right: Navianos de Valverde, Estepa, Sanlúcar de Barrameda, and Casas de Millán. At the center is a frame from a video shared on social media that captures the final part of the luminous phase.} 
    \label{fig:spmn_observations}
\end{figure*}

On the night of Saturday, May 18, 2024, a remarkable superbolide was observed over the Iberian Peninsula at 22:46:48 UTC, specifically crossing Extremadura and northern Portugal before its flight concluded over the Atlantic Ocean. It was recorded by the Spanish Fireball and Meteorite Network \citep[SPMN,][]{2004EM&P...95..553T}, as well as by AMS82, one of the European Space Agency's (ESA) Planetary Defense Office meteor stations of the AllSky7 Network\footnote{\url{www.allsky7.com}}, located in Casas de Millán, Cáceres, Spain. Designated as SPMN180524F by the SPMN network and popularized as the 2024 Iberian Superbolide, this event was notable for its intense brightness which momentarily turned night into day, causing a massive media impact. The superbolide exhibited an atmospherically originated bluish-green glow, leaving a persistent luminous trail in the sky. Table~\ref{tab:stations_det} lists the stations used in this study, while Figure~\ref{fig:spmn_observations} shows the max-combined video frames observed from each station.

Many casual videos immediately surfaced in the media, underscoring the importance of promptly explaining these unusual phenomena to the public. Some of the videos are particularly extraordinary, as the bolide was captured from close distances, revealing significant variations in brightness along its extensive luminous trajectory. These fluctuations in luminosity are typically associated with the continuous fragmentation of the meteoroid and the release of dust and fragments ablated by the heat generated in the frontal shock wave, a characteristic behavior of impacting crumbly meteoroids \citep{2002ESASP.500..233R}. In some videos of the final part of the luminous phase, the bolide head is followed by distinct ablating fragments (see Fig.~\ref{fig:spmn_observations}). 

\begin{table}
\centering
\caption{Longitude, latitude, and altitude of the 4 selected stations recording the SPMN180524F superbolide.}
\label{tab:stations_det}
\begin{tabular}{lccc}
\hline
Station & Lon. ($^\circ$) & Lat. ($^\circ$) & Alt. (m) \\
\hline
Navianos de Valverde	 & 5$^\circ$ 48$'$ 48.4$''$ W & 41$^\circ$ 57$'$ 11.6$''$ N & 711 \\
Estepa & 4$^\circ$ 52$'$ 35.6$''$ W & 37$^\circ$ 17$'$ 29.0$''$ N & 537 \\
Sanlúcar de Barrameda & 6$^\circ$ 20$'$ 31.0$''$ W & 36$^\circ$ 46$'$ 30.3$''$ N & 6 \\
Casas de Millán & 6$^\circ$ 19$'$ 43.0$''$ W & 39$^\circ$ 48$'$ 57.3$''$ N & 456 \\
\hline
\end{tabular}
\end{table}

We calibrate the cameras using the \textit{SkyFit} program provided in the \textit{RMS} library, which incorporates a 7th-order polynomial distortion model and accounts for atmospheric refraction \citep{Vida2021MNRAS5065046V}. We use our \textit{3D-FireTOC} software to analyze the event. This tool integrates the plane intersection method for trajectory triangulation and an N-body numerical integrator for heliocentric orbit determination \citep{2021MNRAS.504.4829P, 2021AsDyn...5..347P, 2023MNRAS.520.5173P, 2023P&SS..23805802P}. Using a model of atmospheric mass density and the measured velocity, the dynamic strength where the meteoroid disruption occurs is computed as $S=\rho_{atm}v^2$ \citep{1983pmp..book.....B}. A traditional method for classifying fireballs is the $P_E$ criterion established by \citet{Ceplecha1976JGR}, which serves to evaluate physical properties of meteoroids. Inspired by this criterion, \citet{2022A&A...667A.158B} introduced a new parameter for assessing the impactor strength, based on the maximum dynamic pressure for large meteoroids. This parameter, termed the pressure resistance factor or pressure factor, is defined as follows:

\begin{equation}
Pf = 100 S_{max} \sin{\overline{\gamma}}^{-1}  m_{0}^{-1/3}  v_{0}^{-3/2},
\label{eq:Pf}
\end{equation}

where \( S_{max} \) represents the maximal dynamic pressure in MPa, $\overline{\gamma}$ is the average slope of the trajectory from the local horizon, \( m_0 \) denotes the initial photometric mass in kg, and \( v_{0} \) is the entry velocity in km\,s\(^{-1}\).

At the start of the luminous phase and down to altitudes near 60 km, the atmosphere is not dense enough to cause observable deceleration. Consequently, all measured points during this phase equally represent the initial velocity, allowing for a robust estimate despite the point-by-point measurement dispersion.


\subsection{Space-based observation}

\begin{figure}
	\includegraphics[width=\columnwidth]{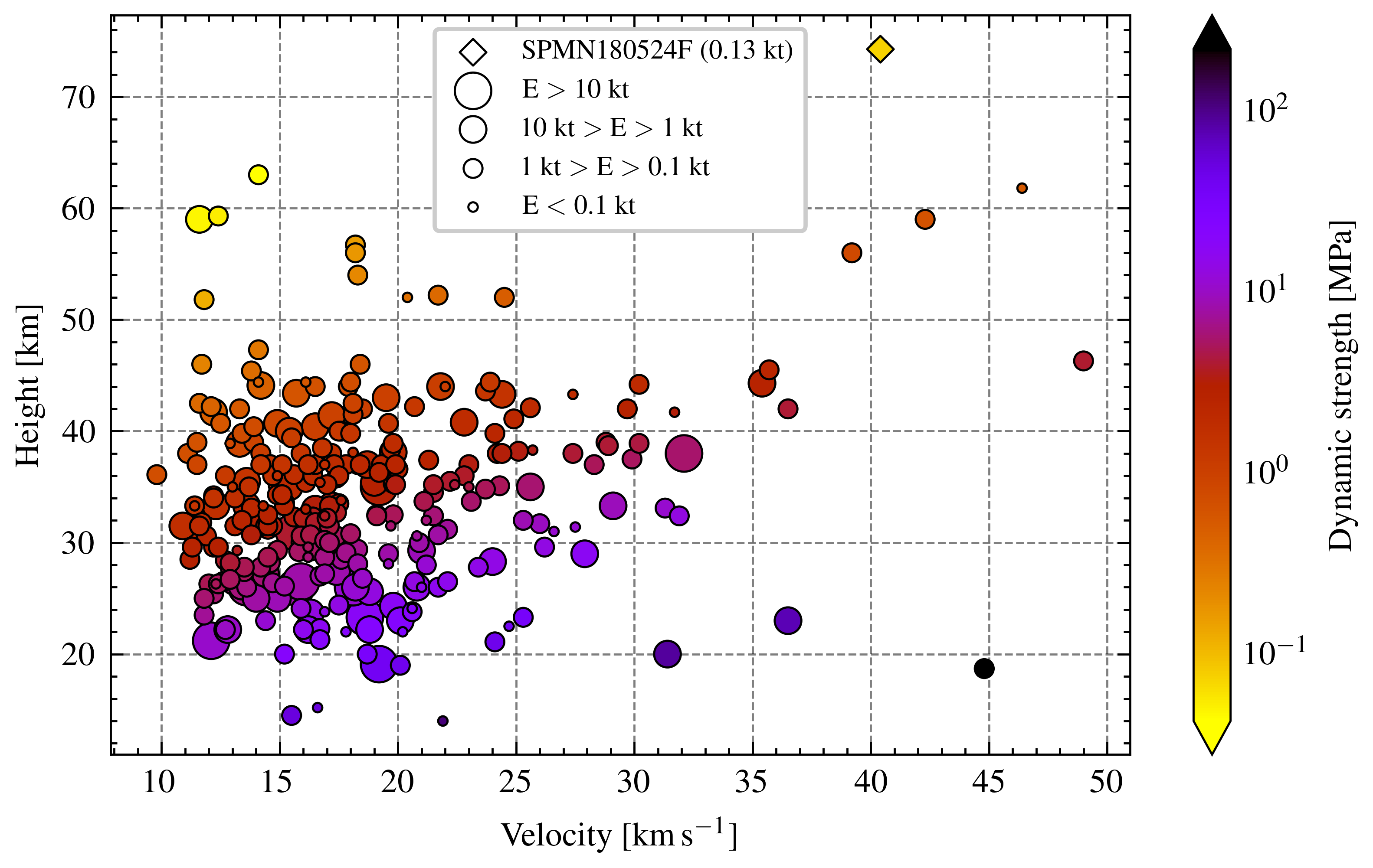}
    \caption{CNEOS fireballs with sufficient data (velocity and height at the peak of radiated energy) to estimate the projectile dynamic strength.}
    \label{fig:cneos_vel_alt_rho}
\end{figure}

From the perspective of the CNEOS database, the SPMN180524F superbolide, with a total impact energy of 0.13 kt, is not a common event: it holds the highest altitude in the database, ranks as the fifth fastest with respect to the Earth, and is the fourth with the lowest dynamic strength (see Fig.~\ref{fig:cneos_vel_alt_rho}). We estimated the CNEOS  fireball density in the same way as described in the previous section.

As the superbolide did not penetrate deeply into the atmosphere and was observed from long distances, the ground-based measurements are insufficient for fitting a dynamic deceleration model. Therefore, we use as a proxy the total impact energy ($E$) measured by the USG sensors, which has proven reliable. By assuming that the energy recorded is equal to the kinetic energy, we estimate the initial mass of the projectile as:

\begin{equation}
m_0 = \frac{2 E}{v^2}	.\label{eq:mass}
\end{equation}

By levering the dynamic, dimensionless approach based on the single body theory, the so-called $\alpha$-$\beta$ method (ballistic coefficient and mass-loss parameter, respectively), we can characterize the atmospheric flight by reverting Eq. 14 of \citet{2009AdSpR..44..323G}:

\begin{equation}
\alpha = \frac{c_d \rho_{sl} h_{h} A}{2 m_0^{1/3} \rho_m^{2/3} \sin \overline{\gamma} },
\end{equation}

where $c_d=0.7$ is the drag coefficient, $A=1.21$ is the spherical shape factor, $\rho_{sl}=1.29\,\,$kg$\,$m$^{-3}$ is the atmospheric density at the sea level, $\rho_m$ is the meteoroid density, and $h_{h}=7.16 \,\,$km is the height of the homogeneous atmosphere. By assuming the final mass is equal to zero, $\beta$ can be estimated from the final height $h_e$ and $\alpha$ \citep{2015Icar..250..544M}:

\begin{equation}
\beta = \frac{e^{h_e/h_0}}{2 \alpha}.
\end{equation}

With $\alpha$ and $\beta$ determined, we fit the atmospheric flight model to the observed data to obtain the velocity profile \citep{2007SoSyR..41..509G}:

\begin{equation}
h(v)=\ln 2\alpha +\beta -\ln (\overline{E}i(\beta )-\overline{E}i(\beta v^{2})), \label{eq_y}
\end{equation}

where 
\[
\overline{Ei}(x)=\int_{-\infty }^{x}\frac{e^{z}dz}{z}. 
\]


To search for possible associations with specific parent bodies or meteoroid streams, we use both traditional D-criteria and Machine Learning distance metrics \citep{PeaAsensio2024}. We employ the NEOMOD model to explore the possible source of the heliocentric orbit \citep{Nesvorny2023AJ, Nesvorny2024Icar, Nesvorn2024Icar41716110N}. NEOMOD integrates asteroid orbits from main belt sources and comets, calibrates results with observations from the Catalina Sky Survey, and includes visible albedo information from the Wide-Field Infrared Survey Explorer (WISE).

\section{Results and discussion}

\subsection{Atmospheric flight, physical properties, and orbit}

\begin{figure}
	\includegraphics[width=\columnwidth]{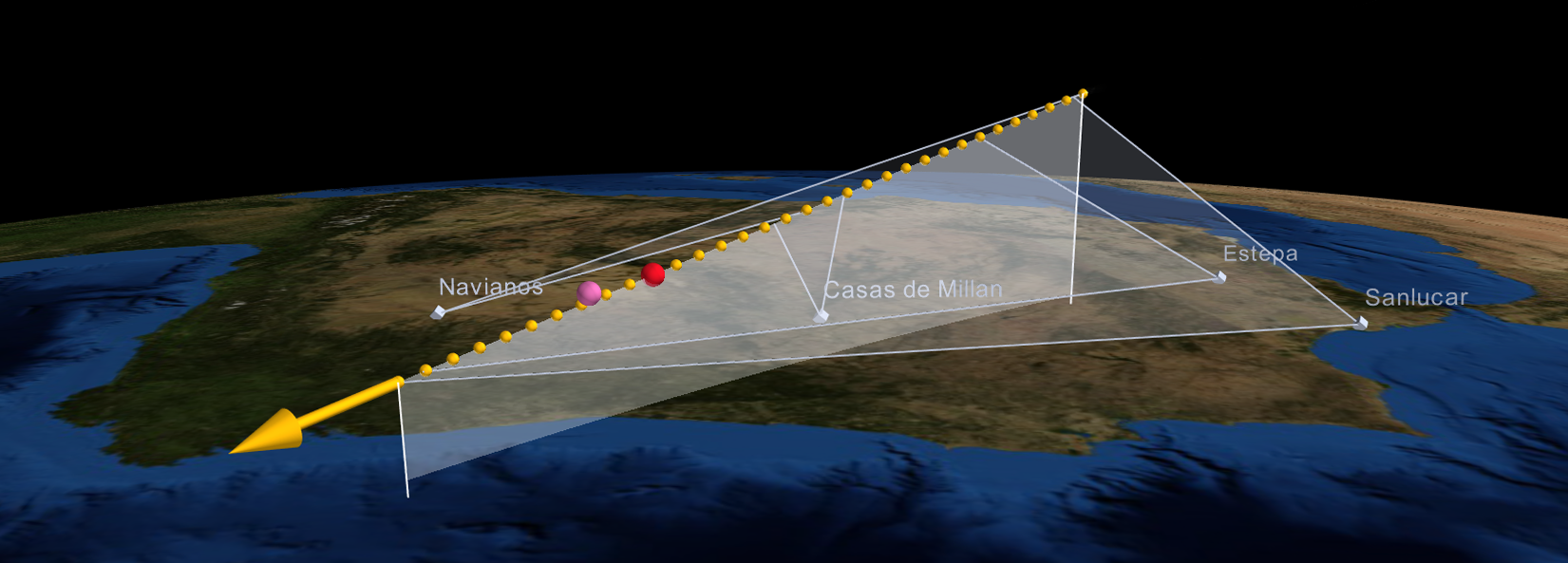}
    \caption{Atmospheric flight of the SPMN180524F superbolide. For ground-based observations, the peak brightness occurs at the red point. In pink is shown the point reported by the USG sensors.}
    \label{fig:3D_reconstruction}
\end{figure}

\begin{figure}
	\includegraphics[width=\columnwidth]{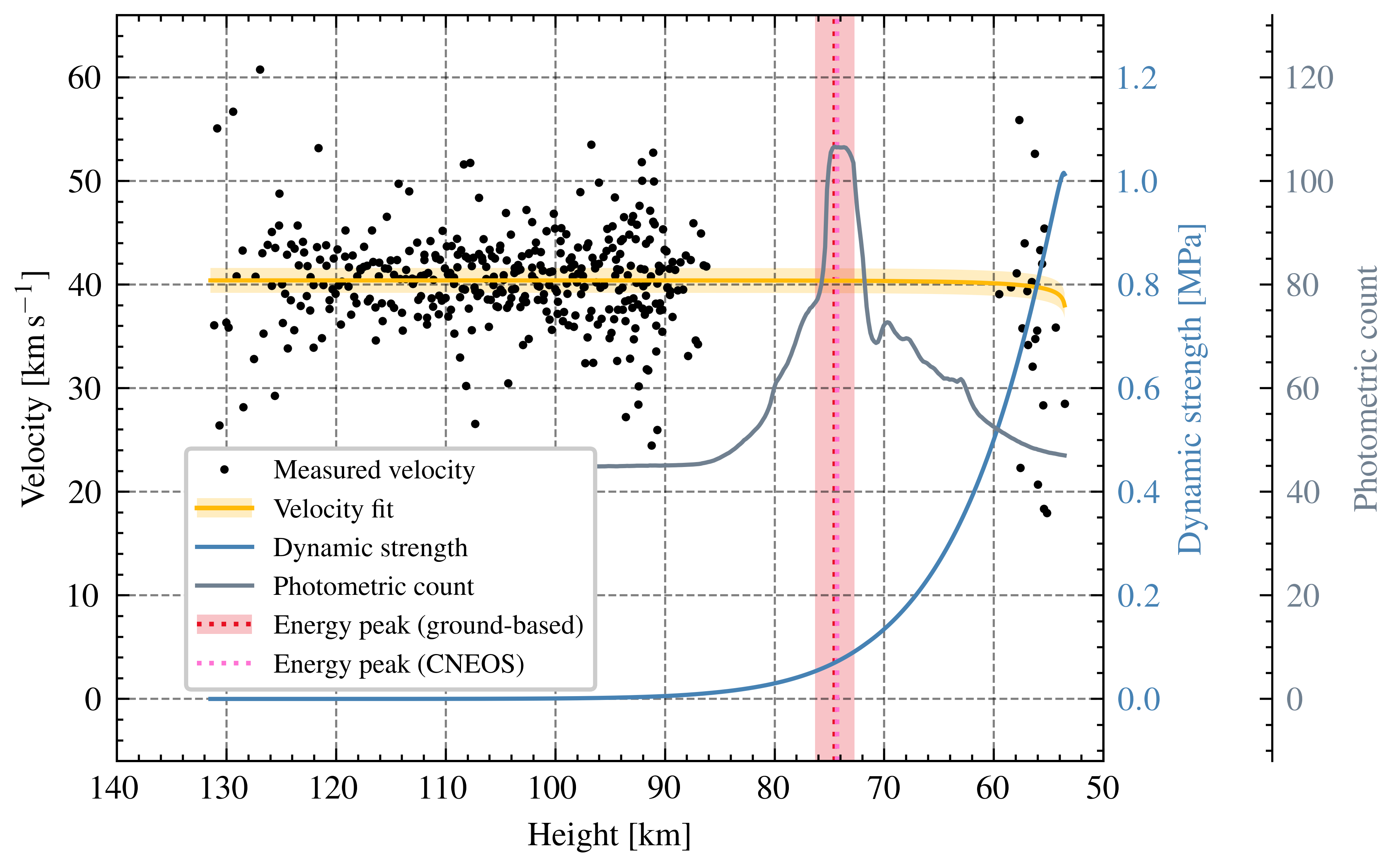}
    \caption{Characterization of the atmospheric flight of the SPMN180524F superbolide. It includes the measured velocity points, the best fit for the velocity with 3-$\sigma$ uncertainty, the dynamic strength, and the uncalibrated photometric counts from Estepa. The section without measured points corresponds to oversaturated frames.}
    \label{fig:vel_alt_S}
\end{figure}

\begin{figure}
	\includegraphics[width=\columnwidth]{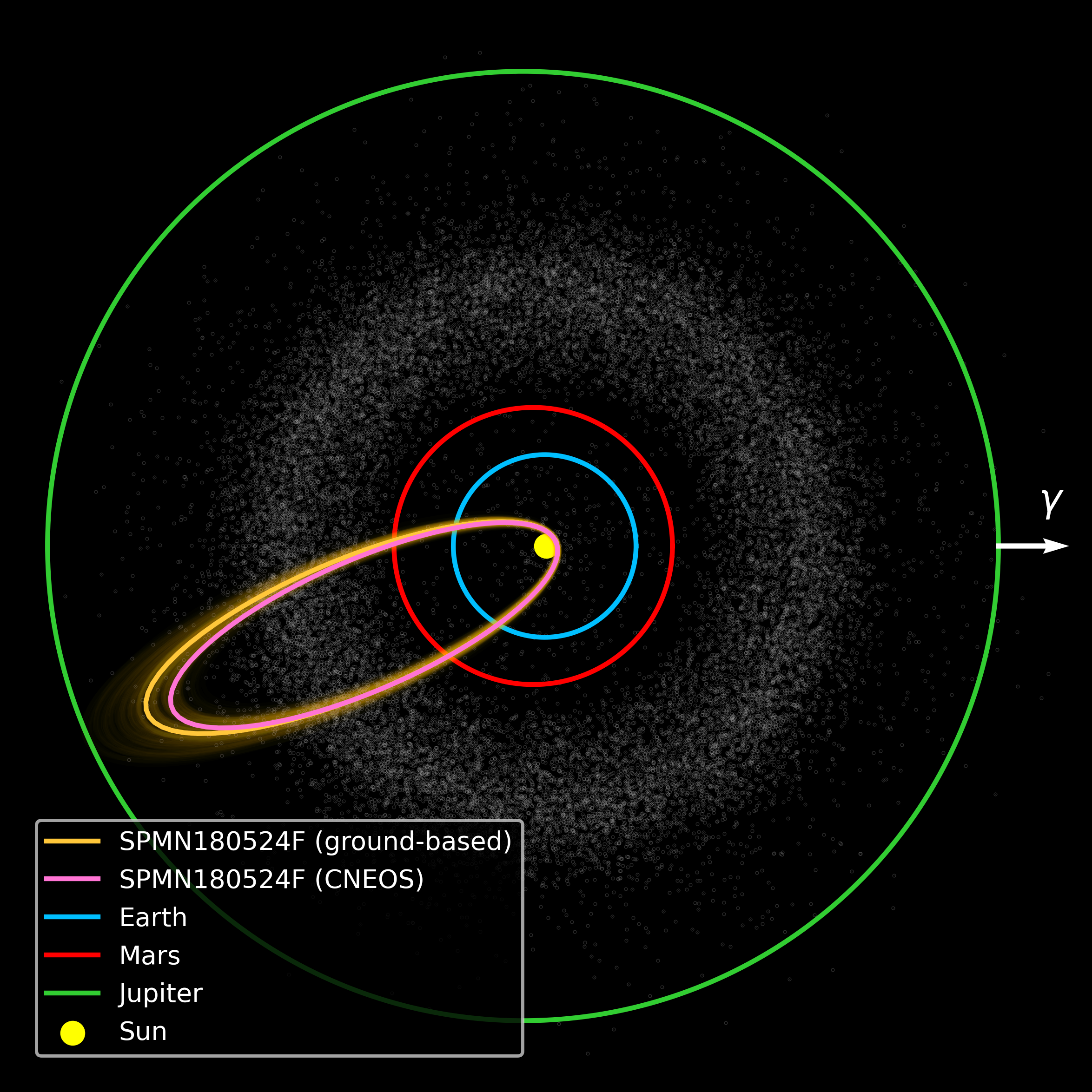}
    \caption{Osculating heliocentric orbit of the SPMN180524F superbolide determined from ground- and space-based observations.}
    \label{fig:orbits}
\end{figure}

Figure~\ref{fig:3D_reconstruction} displays the 3D reconstruction of the atmospheric flight, and Figure~\ref{fig:vel_alt_S} depicts its velocity as a function of height, along with the dynamic strength and the uncalibrated photometric count. The heliocentric orbits derived from ground- and space-based observations are illustrated in Figure~\ref{fig:orbits}. Table~\ref{tab:Results} shows all derived parameters and their comparison where possible.

\begin{table*}
\centering
\footnotesize
\caption{Atmospheric flight, physical parameters, and osculating heliocentric orbital elements of the SPMN180524F superbolide. $^a$Assuming the radiated energy reported by CNEOS is perfectly accurate. $^b$Our value represents the average, whereas CNEOS's value corresponds to the energy peak.
\label{tab:Results}}
\begin{tabular}{lcccc}
\hline
Parameter &  & Ground-based & CNEOS & Discrepancy \\
\hline

Initial time [UTC]                  & $t_0$                         & 2024-05-18 22:46:41   &                       &  \\
Initial longitude [$^\circ$]        & $\lambda_0$                   & 5.4608$\pm$0.0003 W   &                       &  \\
Initial latitude [$^\circ$]         & $\varphi_0$                   & 38.4738$\pm$0.0011 N  &                       &  \\
Initial velocity [km\,s$^{-1}$]     & $v_0$                         & 40.4$\pm$0.4          &                       &  \\
Initial height [km]                 & $h_0$                         & 132.96$\pm$0.16       &                       &  \\

Energy peak time [UTC]              & $t_p$                         & 2024-05-18 22:46:48   & 2024-05-18 22:46:50   & -2 s \\
Energy peak longitude [$^\circ$]    & $\lambda_p$                   & 8.28$\pm$0.11 W       &  8.8 W                & -0.52 \\
Energy peak latitude [$^\circ$]     & $\varphi_p$                   & 40.85$\pm$0.09 N      & 41.0 N                & -0.15 \\
Energy peak velocity [km\,s$^{-1}$] & $v_p$                         & 40.3$\pm$0.4          & 40.4                  & -0.1 \\
Energy peak height [km]             & $h_p$                         & 74.5$\pm$1.8          & 74.3                  & 0.2 \\

Final time [UTC]                    & $t_e$                         & 2024-05-18 22:46:55   &                       &  \\
Final longitude [$^\circ$]          & $\lambda_e$                   & 9.724$\pm$0.006 W     &                       &  \\
Final latitude [$^\circ$]           & $\varphi_e$                   & 41.989$\pm$0.007 N    &                       &  \\
Final velocity [km\,s$^{-1}$]       & $v_e$                         & 38.3$\pm$0.4          &                       &  \\
Final height [km]                   & $h_e$                         & 53.59$\pm$0.24        &                       &  \\

Length [km]                         & $\Delta l$                    & 546.2$\pm$0.9         &                       &  \\
Slope [$^\circ$]                    & $\overline{\gamma}$           & 8.44$\pm$0.05         & 6.5                   &  1.94\textcolor{blue}{$^b$} \\
Azimuth [$^\circ$]                  & $\overline{\phi}$             & 317.05$\pm$0.03       & 315.8                 &  1.25\textcolor{blue}{$^b$}  \\

Dynamic strength [kPa]              & $S$                           & 73$\pm$19             &  72.4                 & 0.6 \\
Meteoroid mass [kg]                 & $m_0$                         & 669$\pm$13\textcolor{blue}{$^a$}        & 667                   &  2 \\
Meteoroid density [kg\,m$^{-3}$]    & $\rho_m$                      & 1500$\pm$300          &                  &  \\
Initial diameter [m]                & $D$                           & 1.0$\pm$0.1           &                   &   \\ 

Geo. velocity [km\,s$^{-1}$]        & $v_R$                         & 38.5$\pm$0.4          & 38.6                  & -0.1 \\
Geo. radiant R.A. [$^\circ$]        & $\alpha_R$                    & 261.79$\pm$0.03       & 262.4                 & -0.61 \\
Geo. radiant Dec. [$^\circ$]        & $\delta_R$                    & -29.11$\pm$0.07       & -29.7                 &  0.59 \\

Semi-major axis [au]                & $a$                           & 2.43$\pm$0.15         & 2.3                   &  0.13 \\
Eccentricity                        & $e$                           & 0.952$\pm$0.004       & 0.95                  &  0.002 \\
Inclination [$^\circ$]              & $i$                           & 16.36$\pm$0.31        & 18.2                  & -1.84 \\
Argument of perihelion [$^\circ$]   & $\omega$                      & 144.64$\pm$0.14       & 145.4                 & -0.76 \\
Long. of the asc. node [$^\circ$]   & $\Omega$                      & 238.1262$\pm$0.0001   & 238.1321              & -0.0059 \\
Perihelion distance [au]            & $q$                           & 0.116$\pm$0.003       & 0.11                  &  0.006 \\
True anomaly [$^\circ$]             & $f$                           & 215.36$\pm$0.14       & 214.6                 &  0.76 \\
Period [year]                       & $P$                           & 3.79$\pm$0.34         & 3.5                   &  0.29 \\
Tisserand's parameter               & $T_J$                         & 2.55$\pm$0.13         & 2.66                  & -0.11 \\
Ballistic coefficient               & $\alpha$                      & 23.6$\pm$3.1          &                       &  \\
Mass-loss parameter                 & $\beta$                       & 38$\pm$5              &                       &  \\
\hline
\end{tabular}
\end{table*}

\begin{figure*}
	\includegraphics[width=2\columnwidth]{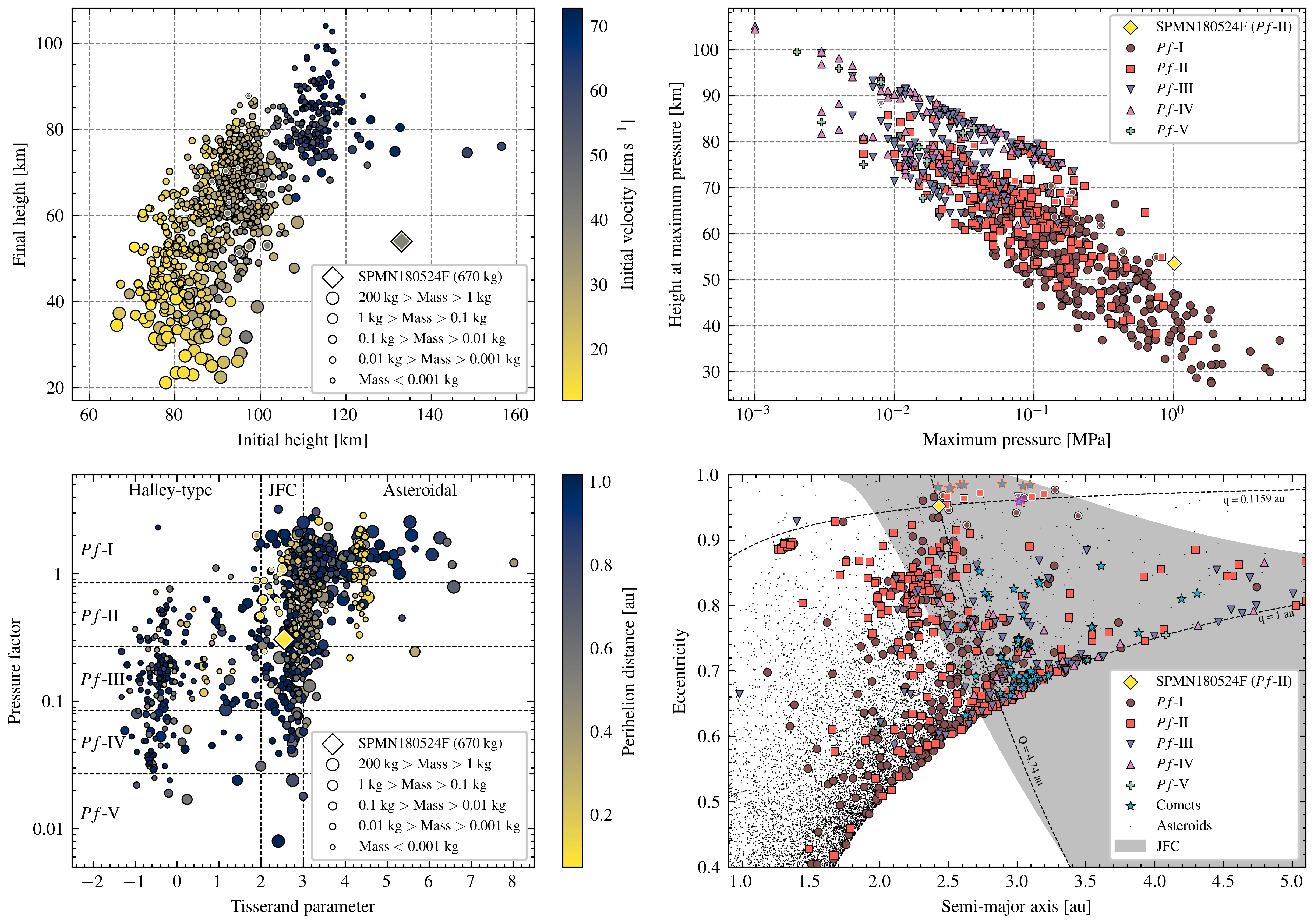}
    \caption{Comparison of the EN catalog and the SPMN180524F superbolide. Top left: final and initial heights, color-coded by initial velocity. Top right: height at maximum pressure and maximum pressure, classified by the pressure factor. Bottom left: pressure factor and Tisserand parameter with respect to Jupiter, color-coded by perihelion distance. Bottom right: eccentricity and semi-major axis distribution, classified by the pressure factor. Asteroids and comets with perihelion distances lower than 1 au are plotted. The inner white edge denotes fireballs that originate with 100\% confidence from a JFC source (see Table~\ref{tab:FN_similar}). Comets with an orange edge represent the SOHO family, and those with a pink edge are from the Machholz complex.}
    \label{fig:context}
\end{figure*}

For the SPMN180524F superbolide, we computed a $P_E$ value of -5.15, which is slightly above the threshold of -5.25 that differentiates between carbonaceous and regular cometary material types. To contextualize this event, we compared it to the catalog of 824 fireballs recorded and analyzed by the European Fireball Network (EN) \citep{2022A&A...667A.157B, 2022A&A...667A.158B}. We calculated a $Pf$ value of 0.31 for SPMN180524F, placing it in class $Pf$-II. Meteoroids with $Pf$ values less than 0.27 are considered cometary, while those with $Pf$ greater than 0.85 are considered asteroidal. Therefore, we interpret SPMN180524F as a weak carbonaceous body and draw its bulk density from a uniform distribution between 1000 and 2000 kg\,m$^{-3}$ for subsequent calculations. Figure~\ref{fig:context} shows the comparison with the EN catalog. The SPMN180524F superbolide appears as an outlier in the final versus initial height panel, possibly due to its grazing atmospheric slope and size, as the largest mass in the catalog is 110 kg. This event is not particularly unusual when considering the combination of the Tisserand parameter and the pressure factor, as similar carbonaceous impactors are found in Jupiter-Family Comet (JFC) orbits. SPMN180524F stands out due to its perihelion distance (0.12 au), which is within the shortest in the catalog. In the eccentricity versus semi-major axis panel, the SPMN180524F event appears grouped with other $Pf$-II and $Pf$-I events, lying well within the JFC region and approaching the so-called excited short-period orbits \citep{2022A&A...667A.158B}. SPMN180524F is the second most pressure-resistant $Pf$-II event. Remarkably, the samples returned by the OSIRIS-REx NASA mission revealed similar bulk density values for the carbonaceous asteroid Bennu \citep{Lauretta2024Bennu}. We note that carbonaceous chondrite (CC) projectiles are difficult to detect before the atmospheric impact due to their typical low albedo \citep{2014MNRAS.437..227T, 2020A&A...641A..58T}

Searching for associations with meteoroid streams or parent bodies yielded negative results. Nevertheless, $\epsilon$-Scorpiids meteor shower exhibits the highest similarity with $D_D$ = 0.08 \citep{Drummond1981Icar...45..545D}. By estimating the source contributions to the major axis, eccentricity, inclination, and absolute magnitude space, NEOMOD determined that the orbit of the SPMN180524F impactor originated from a JFC with a 100\% confidence. On the other hand, while the semi-major axis of $\sim$2.5 au might suggest an origin from the 3:1 resonance, this hypothesis is not supported by NEOMOD and should be verified numerically. We emphasize that the dynamic transit of an asteroid from the outer main asteroid belt cannot be ruled out due to the chaotic nature of these processes \citep{1995Icar..118..169V}. We have also run the NEOMOD model on the EN catalog and listed in Table~\ref{tab:FN_similar} the events with 100\% confidence of having a JFC source. Among these, SPMN180524F was the only meter-sized impactor, had the highest initial altitude by a significant margin and the shallowest atmospheric entry angle, and experienced the greatest dynamic strength. Note that NEOMOD can extrapolate results for two-thirds of the catalog, and only 7.6\% of events with $T_J$ between 2 and 3 have over a 95\% probability of originating from a JFC source.

\begin{table*}
\centering
\caption{List of EN fireballs with orbits that NEOMOD identifies as 100\% originating from JFC sources. The superbolide SPMN180524F is included for comparison. The events are sorted by photometric mass.}
\label{tab:FN_similar}
\begin{tabular}{lccccccccccc}
\hline
Event & $m_0$ & $h_0$ & $h_e$ & $\overline{\gamma}$ & $v_0$ & $S_{max}$ & $Pf$ class & $T_J$ & $q$ & $e$ & $i$ \\
 & [kg] & [km] & [km] & [$^\circ$] & [km\,s$^{-1}$] & [MPa] &  & & [au] & & [$^\circ$] \\
\hline
SPMN180524F      & 670  & 132.96 & 53.59 &       & 40.4 & 1.015 & $Pf$-II & 2.55 & 0.116 & 0.952 & 16.36  \\
EN020617\_230339 & 0.57 & 97.37 & 52.80 & 12.56 & 36.993 & 0.434 & $Pf$-I & 2.511 & 0.2017 & 0.9234 & 11.52  \\
EN020818\_012951 & 0.45 & 101.62 & 53.04 & 39.87 & 41.273 & 0.819 & $Pf$-II & 2.043 & 0.1052 & 0.9661 & 22.67  \\
EN040818\_014525 & 0.035 & 100.45 & 66.95 & 25.24 & 41.862 & 0.177 & $Pf$-II & 1.976 & 0.0937 & 0.9707 & 22.96  \\
EN180517\_013552 & 0.015 & 92.38 & 60.52 & 25.68 & 39.743 & 0.302 & $Pf$-I & 2.501 & 0.1333 & 0.9467 & 19.69  \\
EN040717\_223844 & 0.011 & 95.06 & 51.30 & 38.92 & 41.700 & 0.755 & $Pf$-I & 1.890 & 0.2163 & 0.9372 & 48.15  \\
EN300717\_021036 & 0.0072 & 98.31 & 69.04 & 23.07 & 42.998 & 0.189 & $Pf$-I & 1.890 & 0.0757 & 0.9769 & 27.50  \\
EN160817\_004534 & 0.005 & 88.68 & 67.88 & 32.11 & 42.119 & 0.181 & $Pf$-II & 2.220 & 0.0757 & 0.9722 & 22.47  \\
EN080818\_005515 & 0.0018 & 91.93 & 70.30 & 27.49 & 41.386 & 0.133 & $Pf$-I & 2.068 & 0.1067 & 0.9650 & 25.00  \\
EN180717\_000606 & 0.001 & 101.31 & 79.05 & 16.73 & 41.098 & 0.037 & $Pf$-II & 2.440 & 0.1158 & 0.9534 & 33.60  \\
EN161118\_233946 & 0.00083 & 99.36 & 66.96 & 66.2 & 41.093 & 0.142 & $Pf$-II & 2.365 & 0.0944 & 0.9638 & 8.49  \\
EN061218\_013732 & 0.00052 & 92.64 & 63.04 & 67.39 & 42.235 & 0.203 & $Pf$-I & 2.438 & 0.0782 & 0.9682 & 20.60  \\
EN091018\_014119 & 0.00037 & 100.69 & 63.52 & 51.10 & 37.664 & 0.140 & $Pf$-I & 2.247 & 0.1740 & 0.9419 & 4.61  \\
EN081118\_011907 & 0.00015 & 94.78 & 71.09 & 55.34 & 42.043 & 0.073 & $Pf$-II & 2.409 & 0.0844 & 0.9661 & 26.38  \\
EN060818\_212305 & 0.00013 & 97.26 & 87.75 & 13.11 & 41.819 & 0.008 & $Pf$-III & 2.083 & 0.1007 & 0.9665 & 25.44  \\
\hline
\end{tabular}
\end{table*}

The plausible connection between carbonaceous chondrites (CCs) and comets, long suggested, is further supported by recent studies of the reflectance spectra of comet 2P/Encke and ungrouped CCs \citep{2020A&A...641A..58T}. The role of erosion and subsequent dehydration caused by thermal processing of cometary nuclei was elucidated by Rosetta's study of 67P/Churyumov-Gerasimenko, providing insights into the evolution of meteoroids resulting from their disruption \citep{2020MNRAS.493.4039F, 2019SSRv..215...34K, 2019SSRv..215...18T}. Research on cometary formation and disintegration products suggests that centimeter-sized pebbles with higher density can be preserved in the interior of comets \citep{2022Univ....8..381B, 2022NatAs...6..546C, 2022MNRAS.509.5641S, 2022MNRAS.512.2277T}. However, these materials are too friable to explain the fragments observed trailing the bolide head at aerodynamic pressures of 1 MPa (see Fig.~\ref{fig:cneos_vel_alt_rho}). Even carbonaceous bodies have been determined to suffer atmospheric fragmentation at $\sim$0.5 MPa \citep{2024arXiv240619727B}, with ordinary chondritic surviving greater pressures \citep{Popova2011MPS}. The Almahata Sitta meteorite, a fall associated with the impact of asteroid 2008 TC3, exemplifies a polymict breccia \citep{2010M&PS...45.1638B, 2010M&PS...45.1590J, 2010M&PS...45.1778K}, and recent work suggested that was a polymict C1 chondrite parent body \citep{2022M&PS...57.1339B}. Similarly, we propose that a good scenario for the SPMN180524F superbolide is that the meter-sized weak carbonaceous meteoroid encountered asteroidal debris, becoming a breccia with higher-strength macroscopic particles in its interior. It is known that common cometary outgassing cannot release meter-sized meteoroids, indicating that the SPMN180524F meteoroid may be a remnant of a disruption of its parent comet under the thermal stress imposed by solar heat near its close perihelion \citep{2006mspc.book.....J, 2008tnoc.book....1J, 2022MNRAS.512.2277T}. If the SPMN180524F superbolide resulted from an object disruption, additional fragments could reach the Earth. 

There are examples of superbolides produced by CC projectiles, such as the Tagish Lake meteorite fall, which had an estimated meteoroid density of 1500 kg\,m$^{-3}$ but resulted in higher density ungrouped CCs \citep{2000Sci...290..320B}, as well as the Maribo, Sutter’s Mill, Flensburg, and Winchcombe meteorites \citep{2012M&PS...47...30H, 2012Sci...338.1583J, 2021M&PS...56..425B, 2024M&PS...59..927M}. We also find the polymict breccia hypothesis for the Maribo event plausible, as it had a density of 2000 kg\,m$^{-3}$ at entry, with some fragments surviving 3-5 MPa \citep{2019M&PS...54.1024B}. Table~\ref{tab:carbo_sources} lists the potential orbital sources for these events, as computed by NEOMOD. Despite having Tisserand parameters typical of JFCs (or close to it, as in the case of Winchcombe and Tagish Lake), NEOMOD determined that these events most likely did not originate from a JFC source, as opposed to SPMN180524F. CC impactors on JFC-like orbits generally have high entry velocities due to their eccentricity, resulting in their complete disintegration in the atmosphere. In contrast, CCs on other orbital paths tend to enter the atmosphere at slower velocities, increasing their chances of survival. This leads to a bias in our inventory of recovered CCs. The direct video evidence of the meter-sized CC from the SPMN180524F event, which followed a JFC-like orbit, provides a clear example of such high-velocity atmospheric ablation. For instance, the Winchcombe fireball had an entry velocity of only 13.5 km\,s$^{-1}$, which facilitated fragments reaching the ground. Table~\ref{tab:carbo_sources} also shows two bright fireballs recorded by the SPMN that were tentatively associated with a JFC \citep{2009MNRAS.394..569T, 2022M&PS...57..575H}, as well as the April 13, 2021 bolide observed off the coast of Florida and Grand Bahama Island \citep{2022M&PS...57..575H}. We used a semi-major axis of 4.2 au because NEOMOD cannot extrapolate the results for the nominal values of SPMN110708 and SPMN130413 (instead of 4.5 and 4.3 au, respectively). However, backward numerical integrations over the past 10,000 years suggest that these superbolides are unlikely to be genuine JFCs \citep{Shober2024arXiv240508224S}.


\begin{table}
\centering
\caption{Most likely sources for superbolides produced by carbonaceous or allegedly JFC projectiles as computed by NEOMOD, along with their Tisserand parameters with respect to Jupiter. The estimated probabilities are shown in parentheses.}
\label{tab:carbo_sources}
\begin{tabular}{lcccc}
\hline
Impactor & $T_J$ & 1$^{st}$ source & 2$^{nd}$ source & 3$^{rd}$ source \\
\hline
SPMN180524F & 2.55 & JFC (100\%) & &  \\
SPMN110708 & 2.0 & JFC (99.3\%) & 3:1 (0.4\%) & 2:1 (0.2\%) \\
SPMN130413 & 2.3 & JFC  (97.6\%) & 5:2 (1.1\%) & 3:1 (0.6\%) \\
Apr. 13 2021 & 2.57 & JFC (92.0\%) & 5:2 (5.7\%) & 3:1 (1.3\%) \\
Tagish Lake & 3.52 & $\nu_6$ (78.6\%) & 3:1 (20.2\%) &5:2 (0.6\%) \\
Maribo & 2.95 & 3:1 (75.4\%)& $\nu_6$ (19.6\%)&  5:2 (4.3\%) \\
Sutter’s Mill & 2.81 & 5:2 (53.5\%)& JFC (27.0\%)&  3:1 (13.1\%) \\
Flensburg & 2.89 & 5:2 (81.6\%)& 3:1 (10.6\%)&  JFC (4.1\%) \\
Winchcombe & 3.12 & 3:1 (80.8\%)& $\nu_6$ (14.8\%)&  5:2 (3.7\%) \\
\hline
\end{tabular}
\end{table}

In fact, \citet{Shober2024arXiv240508224S} performed numerical integration and ran NEOMOD on an extensive fireball dataset to demonstrate that the Jupiter-decoupled orbits of fireballs cannot be attributed to a JFC source, contrary to NEOMOD results or the classical Tisserand parameter classification. The study found that most fireballs with JFC-like orbits are decoupled from Jupiter and exhibit more stable orbits. Specifically, 79-92\% of JFC-like meteoroids detected by fireball networks are not subject to frequent Jupiter encounters, and fewer than 5\% of all fireballs exhibit dynamics similar to genuine JFCs, suggesting that the likely source is the main belt. Conversely, all genuine JFCs reside on orbits that frequently encounter Jupiter. The decoupling of impacting meteoroids from Jupiter may result from non-gravitational forces, which NEOMOD does not account for. However, non-gravitational forces affecting meteoroids, including those from JFCs, are generally less significant than those affecting comets. The primary non-gravitational force of interest is the Yarkovsky drift, but this effect is slow and insufficient to cause significant decoupling from Jupiter’s orbit compared to the effects of close encounters with Jupiter \citep{2000Icar..145..301B}. For the specific case of the SPMN180524F superbolide, we computed the Minimum Orbit Intersection Distance (MOID) relative to Jupiter as 1.51 AU using the method described by \citet{2019A&C....27...11B}. This approach yields accurate and rapid results by solving a 16th-order polynomial. This value implies that this impactor with a JFC-like orbit, similar to many identified by \citet{Shober2024arXiv240508224S}, must have been ejected from a body already decoupled from Jupiter’s orbit. The mechanisms responsible for this detachment remain unclear. While secular evolution and non-gravitational forces require extended periods to be effective, frequent close encounters with Jupiter challenge orbital stability. Although the immediate precursor parent body may have evolved from a comet, their orbital dynamics are inconsistent with those of JFCs.

Looking at the bottom-right panel of Figure~\ref{fig:context}, we can observe a cluster of fireballs, including the SPMN180524F superbolide, separated from the rest and close to several comets. These nearby comets are all Solar and Heliospheric Observatory (SOHO) comets, marked with an orange edge. Also in the vicinity, comets of the Machholz complex can be observed, depicted with a pink edge. All of them with extremely stable orbits over 10,000-year timescales, having MOID with Jupiter larger than 0.5 au \citep{Shober2024arXiv240508224S}. These events would have been traditionally classified as near-Earth JFCs, despite not having chaotic behaviors, which is indicative of an evolved population. Whether these objects originated on the main belt, the Kuiper belt, or the scattered disk is an open question. Nonetheless, we note that NEOMOD considers SOHO and Machholz comets as JFC with 100\% confidence. Based on this and following the classification of \citet{2018SSRv..214...20J}, the SPMN180524F superbolide can be termed a sunskirter ($q$ < 0.153 au). 

\citet{2020AJ....159..143W} observed a lack of meter-sized bodies with near-Sun perihelia and an excess of millimeter-sized meteoroids. This suggests that near-Sun objects do not fragment into meter-sized pieces but instead disintegrate into millimeter-sized particles. They propose that the disruption of near-Sun asteroids, along with the brightening and destruction processes affecting SOHO comets, occurs through meteoroid erosion, where material is removed by high-speed near-Sun meteoroid impacts. This idea may align with the polymict breccia hypothesis. However, we cannot rule out the possibility that SPMN180524F was a conventional C-type body that was captured into a stable orbit and subsequently experienced volatile sublimation and cracking. This process could have caused it to resemble a weaker object, with only a few interior fragments surviving the thermal processing, which are the pieces observed at the end of its atmospheric flight. In any case, the SPMN180524F event could provide valuable insights into the supercatastrophic disruption mechanisms of sunskirters \citep{2016Natur.530..303G}. In future studies, we will conduct a numerical analysis of the dynamic evolution of SPMN180524F to evaluate its stability and gain further insights into its origin.


\subsection{Comparison with CNEOS data}

Efforts have been made to estimate the uncertainties of the CNEOS database based on ground-based observations, revealing two groups of measurements: one with sufficient accuracy to allow acceptable heliocentric orbits and another one with significant radiant and velocity deviations. \citet{2019MNRAS.483.5166D} reported discrepancies in the radiants of CNEOS fireballs, ranging from a few degrees to as much as 90$^\circ$. For example, velocity vectors were inaccurately measured for events such as Buzzard Coulee, 2008 TC3, Kalabity, and Crawford Bay. Specific typographical errors, like the missing minus sign in the $z$ velocity component for 2008 TC3, were noted by \citet{2022AJ....164...76P}, in addition to comparing two new events independently measured (2019 MO and 2022 EB5). Further independent analyses have included events like Saricicek, Ozerki, Viñales, Flensburg, Novo Mesto, and Adalen, which helped refine the mean radiant and velocity deviations of CNEOS fireballs \citep{2023ApJ...953..167B, 2024Icar..40815844P}.


The last column of Table~\ref{tab:Results} lists the discrepancy with the values derived from CNEOS data. All the parameters compared were obtained independently, except for mass, which was derived from the radiated energy provided by CNEOS, assuming it is perfectly accurate. The SPMN180524F superbolide belongs to the group of events well-measured from space, as the apparent radiant is deviated in 0.56$^\circ$ and the velocity at the energy peak in 0.1 km\,s$^{-1}$, resulting in a good agreement on the heliocentric orbit. One notable divergence is the -1.84$^\circ$ discrepancy in the orbital inclination. Although the CNEOS detection is very close to the atmospheric trajectory, the energy peak coordinate appears offset by $\sim$60 km, despite matching in height. This discrepancy may result from the inherent error in the USG space sensor observations. Nevertheless, there are also challenges in determining the point of maximum brightness from ground-based stations due to highly saturated frames, variations introduced by camera optics, and observation conditions, particularly for distant detections.

\section{Conclusions}

The 18 May 2024 superbolide was a unique event, demonstrating how a relatively fragile meter-sized meteoroid can produce a spectacular display of color and luminosity. It also exemplified the Earth's atmosphere as an excellent shield for this type of impactor. We reconciled the satellite radiometric data with a purely dynamic atmospheric flight model to constrain the meteoroid’s mass and consistently derive the atmospheric velocity profile. The analysis of its characteristics indicates that the most likely source of the carbonaceous meteoroid is the JFC region, aligning with the SOHO comet family, as its sunskirting orbit is decoupled from Jupiter. To explain the presence of fragments surviving pressures of 1 MPa, we hypothesize that the meteoroid could contain collisionally implanted higher-strength macroscopic particles, forming a polymict carbonaceous chondrite object, or it may be a heavily thermally processed C-type asteroid.


Our analysis demonstrated good agreement with the data reported by CNEOS both in radiant and velocity, and subsequently in the heliocentric orbital elements. Additionally, we compared it with the EN fireball catalog and identified similar, but centimeter-sized, events. The presence of meter-sized objects in the vicinity of the Sun provides new constraints on the timescales and characteristics of supercatastophic disruption mechanisms. From an impact risk perspective, this event raises questions about why such an impactor was not detected by current telescopic surveys, which have successfully identified some asteroids of a few meters before their collision with Earth. Our results provide a clear explanation: the meteoroid was too small and had a low albedo, making it hardly detectable.

\section*{Acknowledgements}

EP-A acknowledges support from the LUMIO project funded by the Agenzia Spaziale Italiana. JMT-R and PG-T acknowledge support from the Spanish project PID2021-128062NB-I00 funded by MCIN/AEI, which sustains the SPMN network. PG-T also acknowledges the FPI predoctoral fellowship PRE2022-104624 from the Spanish MCIU. EP-A and JMT-R conducted this work under the Fundación Seneca project (22069/PI/22), Spain. We also acknowledge the monitoring effort made by the members of the SPMN Network. In particular, Miguel A. Furones, Miguel A. Garcia, and Antonio J. Robles contributed to the recordings analyzed in this work. This work is partly based on data from AMS82 cameras of the AllSky7 network, and we thank the network operators, especially Juan Carlos Martín for providing the data. We thank Denis Vida for his assistance in improving the camera calibrations. We also extend our special thanks to Patrick Shober, William F. Bottke, David Nesvorný, and Rogerio Deienno for their help with the orbit source analysis. We also thank Addi Bischoff, Jürgen Blum, and Giovanni Valsecchi for additional insights.


\section*{Data Availability}

The data underlying this article will be shared on reasonable request to the corresponding author.



\bibliographystyle{mnras}
\bibliography{example} 







\label{lastpage}
\end{document}